# Tetraquarks with charm in coupled-channel formalism.

Gerasyuta S.M.[1,2], Kochkin V.I.[1]

1. Department of Theoretical Physics, St. Petersburg State University, 198904, St. Petersburg, Russia.
2. Department of Physics, LTA, 194021, St. Petersburg, Russia,
   E-mail: gerasyuta@sg6488.spb.edu

Abstract

The relativistic four-quark equations are found in the framework of coupled-channel formalism. The dynamical mixing of the meson-meson states with the four-quark states is considered. The approximate solutions of these equations using the method based on the extraction of leading singularities of the amplitudes are obtained. The four-quark amplitudes of cryptoexotic mesons including the quarks of three flavours (u, d, s) and the charmed quark are constructed. The poles of these amplitudes determine the masses of tetraquarks. The mass values of low-lying tetraquarks are calculated.

# I. Introduction.

Hadron spectroscopy always played an important role in the revealing mechanism underlying the dynamic of strong interactions.

Phenomenological models are expected to provide a scheme for the taking into account the vacuum effects and other effects of a nonperturbative character in the zero-order approximation. Investigations revealed that the spectroscopy of the hadrons is one of the main sources of information about nonperturbative QCD effects.

It is necessary to perform the dynamical calculations of the spectrum of multiquark states. These calculations made it possible to determine the spectrum of the lowest tetraquark states.

The observation of the $X(3872)$ [1-4], the first of the XYZ particles to be seen, brought forward the hope that a multiquark state has been received. Maiani et al. advocate a tetraquark explanation for the $X(3872)$ [5-8]. Belle Collaboration observed the $X(3940)$ in double-charmonium production in the reaction $e^+e^- \to J/\psi + X$.

These states can be considered as indication of the possible existence of the tetraquark states [10-18]. In the review [19] the expectations for the properties of conventional charmonium states and predictions for molecules, tetraquarks and hybrids are described.

In series of papers [20-24] a practical treatment of relativistic three-hadron systems has been developed. The physics of the three-hadron system is usefully described in terms of the pairwise interactions among the three particles. The theory is based on the two principles of unitarity and analyticity, as applied to the two-body subenergy channels. The linear integral equations in a single variable are obtained for the isobar amplitudes. Instead of the quadrature methods of obtaining the set of suitable functions is identified and used as a basis for the expansion of the desired solutions. By this means the coupled integral equations are solved in terms of simple algebra.

In our papers [25-28] relativistic generalization of three-body Faddeev equations was obtained in the form of dispersion relations in the pair energy of two interacting particles. The mass spectra of S-wave baryons including $u$, $d$, $s$, $c$ quarks were calculated by a method based on isolating of the leading singularities in the amplitude.

We searched for the approximate solution of integral three-quark equations by taking into account two-particle and triangle singularities, all the weaker ones being neglected. If we considered such an approximation, which corresponds to taking into account two-body and triangle singularities, and defined all the smooth functions in the middle point the physical region of Dalitz-plot, then the problem was reduced to the one of solving a system of simple algebraic equations.

In our previous papers [29-31] the relativistic five-quark equations for the family of the $\theta$ pentaquarks are constructed. The five-quark amplitudes for the low-lying $\theta$ pentaquarks are constructed. The mass of $\theta^+$ pentaquark with positive parity is found to be smaller than the mass of $\theta^+$ pentaquark with negative parity. Our calculations take into account the contributions to the $\theta$ amplitude not only for the $8-$, $10^*-$plets, but also for the $35-$, $27-$, $10-$plets of $SU(3)_f$.

In the present paper the relativistic four-quark equations are found in the framework of coupled-channel formalism. The dynamical mixing between the meson-meson states and the four-quark states is considered. We examine also the possibility for the hidden and open charm diquark-antidiquark states mixing with the four-quark states. Taking the $X(3872)$ and $X(3940)$ as input we predict the existence of the S-wave tetraquark states with $J^{PC} = 0^{++}, 1^{++}, 2^{++}$ including the $u$, $d$, $s$ -quarks and charmed quark. We believe that the mass spectra for the two approachs in consideration are similar.

The paper is organized as follows. In section II we obtain the relativistic four-particle equations which describe the interaction of the quarks. In section III the coupled equations for the reduced amplitudes are derived. Section IV is denoted to a discussion of the results (Tables I, II) for the mass spectrum of S-wave charmed tetraquarks. In the conclusion, the status of the considered model is discussed.

II. Four-quark amplitudes for the tetraquarks.

We derive the relativistic four-quark equations in the framework of the dispersion relation technique. We use only planar diagrams; the other diagrams due to the rules of $1/N_c$ expansion [32-34] are neglected. The current generates a four-quark system. Their successive pair interactions lead to the diagrams shown in Fig. 1 and Fig. 2. The correct equations for the amplitude are obtained by taking into account all possible subamplitudes. It corresponds to the division of complete system into subsystems with smaller number of particles. Then one should represent a four-particle amplitude as a sum of six subamplitudes:

$$A = A_{12} + A_{13} + A_{14} + A_{23} + A_{24} + A_{34}. \qquad (1)$$

This defines the division of the diagrams into groups according to the certain pair interaction of particles. The total amplitude can be represented graphically as a sum of diagrams. We need to consider only one group of diagrams and the amplitude corresponding to them, for example $A_{12}$. We shall consider the derivation of the relativistic generalization of the Faddeev-Yakubovsky approach for the tetraquark. We shall construct the four-quark amplitude of $c\bar{c}u\bar{u}$ meson, in which the quark amplitudes with the quantum numbers of $0^{-+}$ and $1^{--}$ mesons are included. The set of diagrams associated with the amplitude $A_{12}$ can further be broken down into groups corresponding to subamplitudes: $A_1(s,s_{12},s_{34})$, $A_2(s,s_{23},s_{14})$, $A_3(s,s_{23},s_{123})$, $A_4(s,s_{34},s_{234})$, $A_5(s,s_{12},s_{123})$ (Fig. 1), if we consider the tetraquark with $J^{PC}=2^{++}$. In the case of the description of tetraquark with $J^{PC}=0^{++}$, we need to use seven subamplitudes (Fig. 2).

Here $s_{ik}$ is the two-particle subenergy squared, $s_{ijk}$ corresponds to the energy squared of particles $i$, $j$, $k$ and $s$ is the system total energy squared.

The system of graphical equations is determined by the subamplitudes using the self-consistent method. The coefficients are determined by the permutation of quarks [35, 36].

In order to represent the subamplitudes $A_1(s,s_{12},s_{34})$, $A_2(s,s_{23},s_{14})$, $A_3(s,s_{23},s_{123})$, $A_4(s,s_{34},s_{234})$ and $A_5(s,s_{12},s_{123})$ in the form of a dispersion relation it is necessary to define the amplitudes of quark-quark and quark-antiquark interaction $b_n(s_{ik})$. The pair quarks amplitudes $q\bar{q} \to q\bar{q}$ and $qq \to qq$ are calculated in the framework of the dispersion N/D method with the input four-fermion interaction [37, 38] with the quantum numbers of the gluon [39, 40]. The regularization of the dispersion integral for the D-function is carried out with the cutoff parameters $\Lambda$.

The four-quark interaction is considered as an input:

$$g_v(\bar{q}\lambda I_f \gamma_\mu q)^2 + 2g_v^{(s)}(\bar{q}\lambda\gamma_\mu I_f q)(\bar{s}\lambda\gamma_\mu s) + g_v^{(ss)}(\bar{s}\lambda\gamma_\mu s)^2 \quad (2)$$

Here $I_f$ is the unity matrix in the flavor space (u, d), $\lambda$ are the color Gell-Mann matrices. Dimensional constant of the four-fermion interaction $g_v$, $g_v^{(s)}$ and $g_v^{(ss)}$ are parameters of the model. At $g_v = g_v^{(s)} = g_v^{(ss)}$ the flavor $SU(3)_f$ symmetry occurs. The strange quark violates the flavor $SU(3)_f$ symmetry. In order to avoid additional violation parameters we introduce the scale shift of the dimensional parameters [40]:

$$g = \frac{m^2}{\pi^2}g_v = \frac{(m+m_s)^2}{4\pi^2}g_v^{(s)} = \frac{m_s^2}{\pi^2}g_v^{(ss)}, \quad (3)$$

$$\Lambda = \frac{4\Lambda(ik)}{(m_i + m_k)^2}. \quad (4)$$

Here $m_i$ and $m_k$ are the quark masses in the intermediate state of the quark loop. Dimensionless parameters g and $\Lambda$ are supposed to be constants which are independent on the quark interaction type. The applicability of Eq. (2) is verified by the success of De Rujula-Georgi-Glashow quark model [41], where only the short-range part of Breit potential connected with the gluon exchange is responsible for the mass splitting in hadron multiplets.

We use the results of our relativistic quark model [40] and write down the pair quark amplitudes in the form:

$$b_n(s_{ik}) = \frac{G_n^2(s_{ik})}{1 - B_n(s_{ik})}, \quad (5)$$

$$B_n(s_{ik}) = \int_{(m_i+m_k)^2}^{(m_i+m_k)^2 \Lambda/4} \frac{ds'_{ik}}{\pi} \frac{\rho_n(s'_{ik})G_n^2(s'_{ik})}{s'_{ik} - s_{ik}}. \quad (6)$$

Here $G_n(s_{ik})$ are the quark-quark and quark-antiquark vertex functions (Table III). The vertex functions are determined by the contribution of the crossing channels. The vertex functions satisfy the Fierz relations. All of these vertex functions are generated from $g_v$, $g_v^{(s)}$ and $g_v^{(ss)}$. $B_n(s_{ik})$ and $\rho_n(s_{ik})$ are the Chew-Mandelstam functions with cutoff $\Lambda$ [42] and the phase space, respectively:

$$\rho_n(s_{ik}) = \left( \alpha(n) \frac{s_{ik}}{(m_i + m_k)^2} + \beta(n) + \delta(n) \frac{(m_i - m_k)^2}{s_{ik}} \right) \times$$
$$\times \frac{\sqrt{[s_{ik} - (m_i + m_k)^2][s_{ik} - (m_i - m_k)^2]}}{s_{ik}} ,$$
(7)

The coefficients $\alpha(n)$, $\beta(n)$ and $\delta(n)$ are given in Table IV.

Here n=1 corresponds to a $qq$-pair with $J^P = 0^+$ in the $\bar{3}_c$ color state, n=2 describes a $qq$-pair with $J^P = 1^+$ in the $\bar{3}_c$ color state, n=3 corresponds to a $q\bar{q}$-pairs with $J^{PC} = 0^{-+}$ in the $1_c$ color state, n=4 corresponds to a $q\bar{q}$-pairs with $J^{PC} = 1^{--}$ in the $1_c$ color state, and n=5 defines the $q\bar{q}$-pairs corresponding to mesons with quantum numbers: $J^{PC} = 0^{++}, 1^{++}, 2^{++}$.

In the case in question the interacting quarks do not produce a bound state, therefore the integration in Eqs. (8) - (12) is carried out from the threshold $(m_i + m_k)^2$ to the cutoff $\Lambda(ik)$. The coupled integral equation systems, corresponding to Fig. 1 (the meson state with n=5 and $J^{PC} = 2^{++}$ for the $c\bar{c}u\bar{u}$) can be described as:

$$A_1(s, s_{12}, s_{34}) = \frac{\lambda_1 B_4(s_{12}) B_4(s_{34})}{[1 - B_4(s_{12})][1 - B_4(s_{34})]} + 4\hat{J}_2(s_{12}, s_{34}, 4, 4) A_3(s, s'_{23}, s'_{123}),$$
(8)

$$A_2(s, s_{23}, s_{14}) = \frac{\lambda_2 B_4(s_{23}) B_4(s_{14})}{[1 - B_4(s_{23})][1 - B_4(s_{14})]} + 2\hat{J}_2(s_{23}, s_{14}, 4, 4) A_4(s, s'_{34}, s'_{234}) +$$
$$+ 2\hat{J}_2(s_{23}, s_{14}, 4, 4) A_5(s, s'_{12}, s'_{123})$$
(9)

$$A_3(s, s_{23}, s_{123}) = \frac{\lambda_3 B_5(s_{23})}{1 - B_5(s_{23})} + 2\hat{J}_3(s_{23}, 5) A_1(s, s'_{12}, s'_{34}) +$$
$$+ \hat{J}_1(s_{23}, 5) A_4(s, s'_{34}, s'_{234}) + \hat{J}_1(s_{23}, 5) A_5(s, s'_{12}, s'_{123})$$
(10)

$$A_4(s, s_{34}, s_{234}) = \frac{\lambda_4 B_5(s_{34})}{1 - B_5(s_{34})} + 2\hat{J}_3(s_{34}, 5) A_2(s, s'_{23}, s'_{14}) +$$
$$+ 2\hat{J}_1(s_{34}, 5) A_3(s, s'_{23}, s'_{234})$$
(11)

$$A_5(s, s_{12}, s_{123}) = \frac{\lambda_5 B_5(s_{12})}{1 - B_5(s_{12})} + 2\hat{J}_3(s_{12}, 5) A_2(s, s'_{23}, s'_{14}) +$$
$$+ 2\hat{J}_1(s_{12}, 5) A_3(s, s'_{23}, s'_{123})$$
(12)

were $\lambda_i$ are the current constants. They do not affect the mass spectrum of tetraquarks. We introduce the integral operators:

$$\hat{J}_1(s_{12}, l) = \frac{G_l(s_{12})}{[1 - B_l(s_{12})]} \int\limits_{(m_1+m_2)^2}^{(m_1+m_2)^2 \Lambda/4} \frac{ds'_{12}}{\pi} \frac{G_l(s'_{12})\rho_l(s'_{12})}{s'_{12} - s_{12}} \int\limits_{-1}^{+1} \frac{dz_1}{2}, \qquad (13)$$

$$\hat{J}_2(s_{12}, s_{34}, l, p) = \frac{G_l(s_{12})G_p(s_{34})}{[1 - B_l(s_{12})][1 - B_p(s_{34})]} \times$$
$$\times \int\limits_{(m_1+m_2)^2}^{(m_1+m_2)^2 \Lambda/4} \frac{ds'_{12}}{\pi} \frac{G_l(s'_{12})\rho_l(s'_{12})}{s'_{12} - s_{12}} \int\limits_{(m_3+m_4)^2}^{(m_3+m_4)^2 \Lambda/4} \frac{ds'_{34}}{\pi} \frac{G_p(s'_{34})\rho_p(s'_{34})}{s'_{34} - s_{34}} \int\limits_{-1}^{+1} \frac{dz_3}{2} \int\limits_{-1}^{+1} \frac{dz_4}{2}, \qquad (14)$$

$$\hat{J}_3(s_{12}, l) = \frac{G_l(s_{12}, \widetilde{\Lambda})}{1 - B_l(s_{12}, \widetilde{\Lambda})} \times$$
$$\times \frac{1}{4\pi} \int\limits_{(m_1+m_2)^2}^{(m_1+m_2)^2 \widetilde{\Lambda}/4} \frac{ds'_{12}}{\pi} \frac{G_l(s'_{12}, \widetilde{\Lambda})\rho_l(s'_{12})}{s'_{12} - s_{12}} \int\limits_{-1}^{+1} \frac{dz_1}{2} \int\limits_{-1}^{+1} dz \int\limits_{z_2^-}^{z_2^+} dz_2 \frac{1}{\sqrt{1 - z^2 - z_1^2 - z_2^2 + 2zz_1z_2}}, \qquad (15)$$

were $l, p$ are equal to 1 - 5. Here $m_i$ is a quark mass.

In Eqs. (13) and (15) $z_1$ is the cosine of the angle between the relative momentum of the particles 1 and 2 in the intermediate state and the momentum of the particle 3 in the final state, taken in the c.m. of particles 1 and 2. In Eq. (15) $z$ is the cosine of the angle between the momenta of the particles 3 and 4 in the final state, taken in the c.m. of particles 1 and 2. $z_2$ is the cosine of the angle between the relative momentum of particles 1 and 2 in the intermediate state and the momentum of the particle 4 in the final state, is taken in the c.m. of particles 1 and 2. In Eq. (14): $z_3$ is the cosine of the angle between relative momentum of particles 1 and 2 in the intermediate state and the relative momentum of particles 3 and 4 in the intermediate state, taken in the c.m. of particles 1 and 2. $z_4$ is the cosine of the angle between the relative momentum of the particles 3 and 4 in the intermediate state and that of the momentum of the particle 1 in the intermediate state, taken in the c.m. of particles 3, 4.

We can pass from the integration over the cosines of the angles to the integration over the subenergies [43-45].

Let us extract two-particle singularities in the amplitudes $A_1(s, s_{12}, s_{34})$, $A_2(s, s_{23}, s_{14})$, $A_3(s, s_{23}, s_{123})$, $A_4(s, s_{34}, s_{234})$ and $A_5(s, s_{12}, s_{123})$:

$$A_j(s, s_{ik}, s_{lm}) = \frac{\alpha_j(s, s_{ik}, s_{lm}) B_4(s_{ik}) B_4(s_{lm})}{[1 - B_4(s_{ik})][1 - B_4(s_{lm})]}, \quad j=1, 2, \tag{16}$$

$$A_j(s, s_{ik}, s_{ikl}) = \frac{\alpha_j(s, s_{ik}, s_{ikl}) B_5(s_{ik})}{1 - B_5(s_{ik})}, \quad j=3 \text{ - } 5, \tag{17}$$

We do not extract three- particle singularities, because they are weaker than two-particle singularities.

We used the classification of singularities, which was proposed in paper [46]. The construction of approximate solution of Eqs. (8) - (12) is based on the extraction of the leading singularities of the amplitudes. The main singularities in $s_{ik} \approx (m_i + m_k)^2$ are from pair rescattering of the particles i and k. First of all there are threshold square-root singularities. Also possible are pole singularities which correspond to the bound states. The diagrams of Fig.1 apart from two-particle singularities have the triangular singularities, the singularities defining the interaction of four particles. Such classification allows us to search the corresponding solution of Eqs. (8) - (12) by taking into account some definite number of leading singularities and neglecting all the weaker ones. We consider the approximation which defines two-particle, triangle and four-particle singularities. The functions $\alpha_1(s, s_{12}, s_{34})$, $\alpha_2(s, s_{23}, s_{14})$, $\alpha_3(s, s_{23}, s_{123})$, $\alpha_4(s, s_{34}, s_{234})$ and $\alpha_5(s, s_{12}, s_{123})$ are the smooth functions of $s_{ik}$, $s_{ikl}$, $s$ as compared with the singular part of the amplitudes, hence they can be expanded in a series in the singularity point and only the first term of this series should be employed further. Using this classification, one defines the reduced amplitudes $\alpha_1$, $\alpha_2$, $\alpha_3$, $\alpha_4$, $\alpha_5$ as well as the B-functions in the middle point of the physical region of Dalitz-plot at the point $s_0$:

$$s_0^{ik} = 0.25(m_i + m_k)^2 s_0 \tag{18}$$

$$s_{123} = 0.25 s_0 \sum_{\substack{i,k=1 \\ i \neq k}}^{3} (m_i + m_k)^2 - \sum_{i=1}^{3} m_i^2, \quad s_0 = \frac{s + 2\sum_{i=1}^{4} m_i^2}{0.25 \sum_{\substack{i,k=1 \\ i \neq k}}^{4} (m_i + m_k)^2}$$

Such a choice of point $s_0$ allows us to replace the integral Eqs. (8) - (12) (Fig. 1) by the algebraic equations (19) - (23) respectively:

$$\alpha_1 = \lambda_1 + 4\alpha_3 J B_1(4,4,5), \tag{19}$$

$$\alpha_2 = \lambda_2 + 2\alpha_4 JB_2(4,4,5) + 2\alpha_5 JB_3(4,4,5), \tag{20}$$

$$\alpha_3 = \lambda_3 + 2\alpha_1 JC_1(5,4,4) + \alpha_4 JA_1(5) + \alpha_5 JA_2(5), \tag{21}$$

$$\alpha_4 = \lambda_4 + 2\alpha_2 JC_2(5,4,4) + 2\alpha_3 JA_3(5), \tag{22}$$

$$\alpha_5 = \lambda_5 + 2\alpha_2 JC_3(5,4,4) + 2\alpha_3 JA_4(5). \tag{23}$$

We use the functions $JA_i(l)$, $JB_i(l,p,r)$, $JC_i(l,p,r)$ ($l,p,r = 1$ - $5$), which are determined by the various $s_0^{ik}$ (Eq.18). These functions are similar to the functions:

$$JA_4(l) = \frac{G_l^2(s_0^{12})B_l(s_0^{23})}{B_l(s_0^{12})} \int_{(m_1+m_2)^2}^{(m_1+m_2)^2 \Lambda/4} \frac{ds_{12}'}{\pi} \frac{\rho_l(s_{12}')}{s_{12}' - s_0^{12}} \int_{-1}^{+1} \frac{dz_1}{2} \frac{1}{1 - B_l(s_{23}')}, \tag{24}$$

$$JB_1(l,p,r) = \frac{G_l^2(s_0^{12})G_p^2(s_0^{34})B_r(s_0^{23})}{B_l(s_0^{12})B_p(s_0^{34})} \times$$
$$\times \int_{(m_1+m_2)^2}^{(m_1+m_2)^2 \Lambda/4} \frac{ds_{12}'}{\pi} \frac{\rho_l(s_{12}')}{s_{12}' - s_0^{12}} \int_{(m_3+m_4)^2}^{(m_3+m_4)^2 \Lambda/4} \frac{ds_{34}'}{\pi} \frac{\rho_p(s_{34}')}{s_{34}' - s_0^{34}} \int_{-1}^{+1} \frac{dz_3}{2} \int_{-1}^{+1} \frac{dz_4}{2} \frac{1}{1 - B_r(s_{23}')} \tag{25}$$

$$JC_3(l,p,r) = \frac{G_l^2(s_0^{12}, \widetilde{\Lambda})B_p(s_0^{23})B_r(s_0^{14})}{1 - B_l(s_0^{12}, \widetilde{\Lambda})} \frac{1 - B_l(s_0^{12})}{B_l(s_0^{12})} \times$$
$$\times \frac{1}{4\pi} \int_{(m_1+m_2)^2}^{(m_1+m_2)^2 \widetilde{\Lambda}/4} \frac{ds_{12}'}{\pi} \frac{\rho_l(s_{12}')}{s_{12}' - s_0^{12}} \int_{-1}^{+1} \frac{dz_1}{2} \int_{-1}^{+1} dz \int_{z_2^-}^{z_2^+} dz_2 \frac{1}{\sqrt{1 - z^2 - z_1^2 - z_2^2 + 2zz_1z_2}} \times \tag{26}$$
$$\times \frac{1}{[1 - B_p(s_{23}')][1 - B_r(s_{14}')]}$$

$$\widetilde{\Lambda}(ik) = \begin{cases} \Lambda(ik), & \text{if } \Lambda(ik) \leq (\sqrt{s_{123}} + m_3)^2 \\ (\sqrt{s_{123}} + m_3)^2, & \text{if } \Lambda(ik) > (\sqrt{s_{123}} + m_3)^2 \end{cases} \tag{27}$$

The other choices of point $s_0$ do not change essentially the contributions of $\alpha_1$, $\alpha_2$, $\alpha_3$, $\alpha_4$ and $\alpha_5$, therefore we omit the indices $s_0^{ik}$. Since the vertex functions depend only slightly on energy it is possible to treat them as constants in our approximation.

The solutions of the system of equations are considered as:

$$\alpha_i(s) = F_i(s, \lambda_i) / D(s), \tag{28}$$

where zeros of $D(s)$ determinants define the masses of bound states of tetraquarks. $F_i(s, \lambda_i)$ determine the contributions of subamplitudes for the tetraquark amplitude.

III. Calculation results.

The pole of the reduced amplitudes $\alpha_1$, $\alpha_2$, $\alpha_3$, $\alpha_4$, $\alpha_5$ corresponds to the bound state and determines the mass of the meson-meson state with n=5 and $J^{PC}=2^{++}$ for the $c\bar{c}u\bar{u}$ (Fig. 1). If we concider the meson-meson state with $J^{PC}=0^{++}$ (Fig. 2), we must take into account the interaction of the quark and antiquark in the $0^{-+}$ and $1^{--}$ states. In our paper we consider also the interaction of the quarks in the $0^+$ and $1^+$ states that allows us to obtain and compare the diquark-antidiquark states mass spectra (Figs. 3, 4) with the meson-meson state ones (Fig. 1, 2). The quark masses of model $m_{u,d}$=385 MeV, $m_s$=510 MeV coincide with the ordinary meson ones [26, 27] in our model. In order to fix anyhow $m_c$=1586 MeV, we use the tetraquark mass for the $J^{PC}=2^{++}$ $X(3940)$.

The model in consideration has only two parameters:
1) in the case of meson-meson states: the cutoff $\Lambda$=10, and gluon coupling constant $g$=0.794;
2) in the case of diquark-antidiquark states: the cut-off $\Lambda$=10.3, and gluon coupling constant $g$=0.856.

In the first case the parameters are determined by fixing the tetraquark masses for the $J^{PC}=1^{++}$ $X(3872)$ and $J^{PC}=2^{++}$ $X(3940)$ [47]. In the second case we use the tetraquark mass of the $J^{PC}=2^{++}$ $X(3940)$ and the calculated tetraquark mass of the $J^{PC}=0^{++}$ $X(3708)$. The masses of meson-meson states and diquark-antidiquark states with isospin I=0, 1/2, 1 and spin-parity $J^{PC}=0^{++},1^{++},2^{++}$ are predicted (Tables I, II). We use the relativistic four-body approach, which take into account the dynamical mixing of the meson-meson states or diquark-antidiquark states with the four-quark states. We predict the coincidence of the mass spectra of meson-meson states and diquark-antidiquark ones. We believe that the error about 5 per cent is determined by the approximation of the $u$ and $d$ quark interactions in the diquark channel.

In the quark potential models one use the potentials $V_{ud} = \frac{3}{4}V_{0^+} + \frac{1}{4}V_{1^+}$ and $V_{uu} = V_{1^+}$. We can not construct the C-parity for the wavefunctions with the $J^P=1^+$, that the diquark-antidiquark amplitudes for the $|A\rangle = |0_{cq}, 1_{\bar{c}\bar{q}}; 1\rangle$ and $|B\rangle = |1_{cq}, 0_{\bar{c}\bar{q}}; 1\rangle$ states are determined by the similar equations. Therefore we do not consider the diquark-antidiquark states with the $J^P = 1^+$.

## IV. Conclusion.

In a strongly bound system of light and heavy quarks such as the tetraquarks, where p/m~1 for the light quarks, the approximation of nonrelativistic kinematics and dynamics is not justified. We can consider the mixing of ordinary meson masses and the tetraquark ones. In the previous paper [48] we have considered the low-lying charmed mesons as ($u\bar{c}$) and ($s\bar{c}$) states with $J^{PC} = 0^{++}$ and calculated the masses for the $D_0^*$ ($M$ =2109 MeV) and $D_{sJ}^*$ ($M$ =2215 MeV) mesons. In the present paper we have calculated the masses of tetraquarks ($u\bar{c}$)($u\bar{d}$) and ($s\bar{c}$)($u\bar{d}$) with $J^{PC} = 0^{++}$ $M$ ($D_0^*$ tetra) = 2610 MeV and $M$ ($D_{sJ}^*$ tetra) = 2691 MeV. We can consider the mix between ($q\bar{c}$) and ($q\bar{c}$)($q\bar{q}$) states and obtain the experimental values of masses $D_0^*$(2352) and $D_{sJ}^*$(2317) [47].

The contribution of tetraquark with the mass $M$ (2610) to the ordinary meson ($u\bar{c}$) with the mass $M$ (2109) is equal to 10 per sent. The similar contribution of tetraquark with the mass $M$ (2691) to the ordinary meson ($s\bar{c}$) with the mass $M$ (2215) is about 5 per sent. We have shown that the contribution of the tetraquark to the ordinary meson with $J^P = 0^+$ are very small, therefore the width of this state is determined by ordinary state one. The mixing of the ($q\bar{q}$) and ($qq\bar{q}\bar{q}$) states was discussed in the some papers [17, 49], but the contribution of tetraquarks was predicted up to 30 per cent.

Ebert et al. [50, 51] have used the diquark-antidiquark approximation and calculated the tetraquark mass spectrum.The authors have considered the tetraquark as purely the diquark-antidiquark bound states without the contributions of the four-quark states. In distinction we can not use this approach that we obtain the coupled-channel integral equations. Maiani et al. have considered a hyperfine interaction between all quarks and take into account the distinction of the $u$ and $d$ quark masses [5, 6]. The authors predict that the masses of tetraquarks with $J^{PC} = 1^{++}$ and $1^{+-}$ are different and the isospin invariance is broken for the $(Qq)(\overline{Q}\overline{q})$ state in the strong decays. In our model we neglect with the mass distinction of $u$ and $d$ quarks.

We use only two parameters: the cutoff and gluon coupling constant in the distinction of Ebert et al. The authors consider the diquark form factor using the function with the two parameters [50]. In summary, we calculated the masses of heavy tetraquarks with hidden and open charm in the coupled-channel formalism. In contrast to the previous phenomenological consideration of meson-meson molecules with the $\pi$-meson exchange (the mesons inside the molecule are weakly bound) we have shown that the meson-meson states with the gluon exchange interaction possess the mass spectra which are similar to the diquark-antidiquark state ones. We have considered a relativistic four-body problem and investigated the scattering matrix singularities in the coupled-channel formalism. We have determined the contribution of the tetraquark to the ordinary meson with $J^P = 0^+$ which is about 5-10 per sent as compared to ordinary $q\overline{Q}$ state. The interesting opinions with the S-matrix singularities in ref. [52] are given. The approach have developed in this study can be employed to compute the mass spectrum of charmed meson-meson states with open exotica.

## Acknowledgments

The authors would like to thank T. Barnes, S. Capstick and S.L. Zhu for useful discussions. The work was carried with the support of the Russian Ministry of Education (grant 2.1.1.68.26).

Table I. Low-lying meson-meson state masses (MeV).

Parameters of model: quark mass $m_u = 385 MeV$, $m_s = 510 MeV$, $m_c = 1586 MeV$, cutoff parameter $\Lambda = 10$, gluon coupling constant $g = 0.794$.

| Tetraquark | $J^{PC} = 2^{++}$ | $J^{PC} = 1^{++}$ | $J^{PC} = 0^{++}$ |
|---|---|---|---|
| $(c\bar{u})(u\bar{d})$ | 2736 | 2672 | 2610 |
| $(c\bar{s})(u\bar{u})$ $(c\bar{u})(u\bar{s})$ | 2851 | 2770 | 2691 |
| $(c\bar{u})(s\bar{s})$ $(c\bar{s})(s\bar{u})$ | 2975 | 2890 | 2805 |
| $(c\bar{c})(u\bar{u})$ $(u\bar{c})(c\bar{u})$ | 3940 | 3872 | 3708 |
| $(c\bar{c})(s\bar{s})$ $(s\bar{c})(c\bar{s})$ | 4160 | 4020 | 3850 |

Table II. Low-lying diquark-antidiquark state masses (MeV).

Parameters of model: quark mass $m_u = 385 MeV$, $m_s = 510 MeV$, $m_c = 1586 MeV$, cutoff parameter $\Lambda = 10.3$, gluon coupling constant $g = 0.856$.

| Tetraquark | $J^{PC} = 2^{++}$ | $J^{PC} = 0^{++}$ |
|---|---|---|
| $(cu)(\bar{u}\bar{d})$ | 2704 | 2495 |
| $(cu)(\bar{u}\bar{s})$ | 2854 | 2616 |
| $(cs)(\bar{u}\bar{s})$ | 2957 | 2654 |
| $(cu)(\bar{u}\bar{c})$ | 3940 | 3708 |
| $(cs)(\bar{s}\bar{c})$ | 4120 | 3795 |

Table III. Vertex functions.

| $J^{PC}$ | $G_n^2$ |
|---|---|
| $0^+$ (n=1) | $4g/3 - 2g(m_i + m_k)^2/(3s_{ik})$ |
| $1^+$ (n=2) | $2g/3$ |
| $0^{-+}$ (n=3) | $8g/3 - 4g(m_i + m_k)^2/(3s_{ik})$ |
| $1^{--}$ (n=4) | $4g/3$ |
| $2^{++}$ (n=5) | $4g/3$ |
| $1^{++}$ (n=5) | $4g/3$ |
| $0^{++}$ (n=5) | $8g/3$ |

Table IV. Coefficients of Chew-Mandelstam functions.

| $J^{PC}$ | $\alpha$ | $\beta$ | $\delta$ |
|---|---|---|---|
| $0^+$ (n=1) | 1/2 | -e/2 | 0 |
| $1^+$ (n=2) | 1/3 | 1/6-e/3 | -1/6 |
| $0^{-+}$ (n=3) | 1/2 | -e/2 | 0 |
| $1^{--}$ (n=4) | 1/3 | 1/6-e/3 | -1/6 |
| $2^{++}$ (n=5) | 3/10 | 1/5-3e/10 | -1/5 |
| $1^{++}$ (n=5) | 1/2 | -e/2 | 0 |
| $0^{++}$ (n=5) | 1/2 | -1/2 | 0 |

$e = (m_i - m_k)^2 / (m_i + m_k)^2$

Figure captions.

Fig.1. Graphic representation of the equations for the four-quark subamplitudes $A_k$ ($k$=1-5) in the case of n=5 and $J^{PC} = 2^{++}, 1^{++}$ ($c\bar{c}u\bar{u}$).

Fig.2. Graphic representation of the equations for the four-quark subamplitudes $A_k$ ($k$=1-7) in the case of n=5 and $J^{PC} = 0^{++}$ ($c\bar{c}u\bar{u}$).

Fig.3. Graphic representation of the equations for the four-quark subamplitudes $A_k$ ($k$=1-4) in the case of n=5 and $J^{PC} = 2^{++}$ ($\bar{c}\bar{u}cu$).

Fig.4. Graphic representation of the equations for the four-quark subamplitudes $A_k$ ($k$=1-5) in the case of n=5 and $J^{PC} = 0^{++}$ ($\bar{c}\bar{u}cu$).

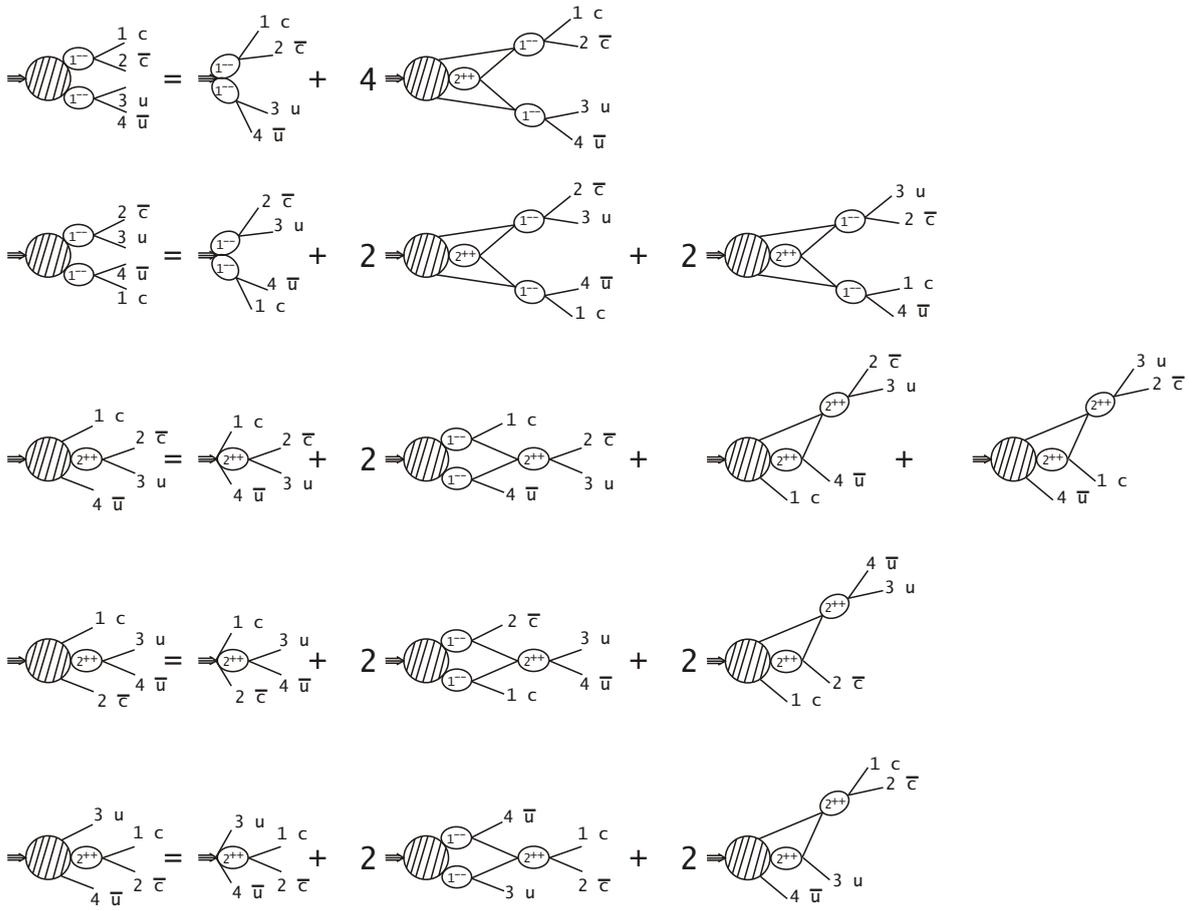

Fig. 1

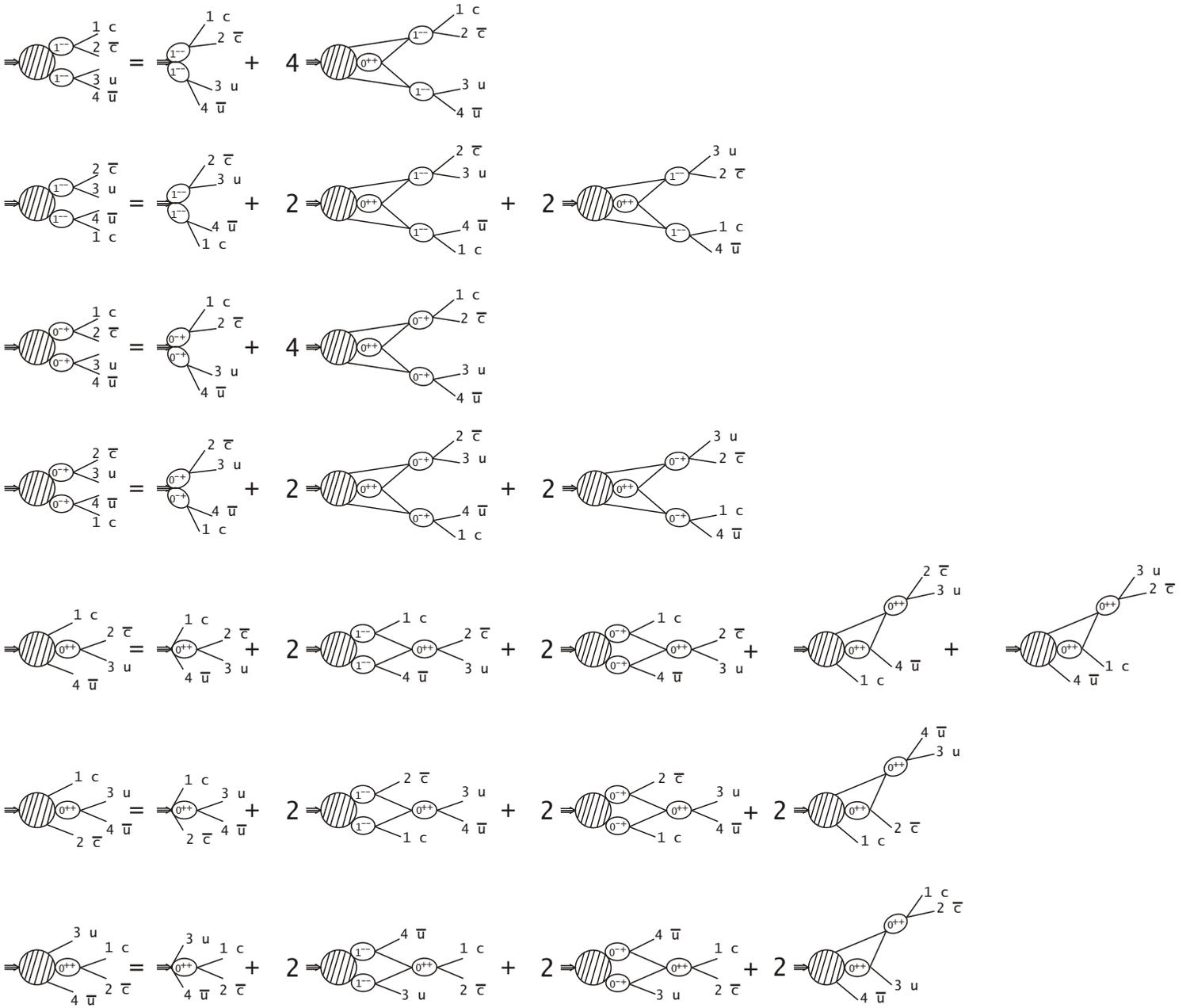

Fig. 2

Fig. 3

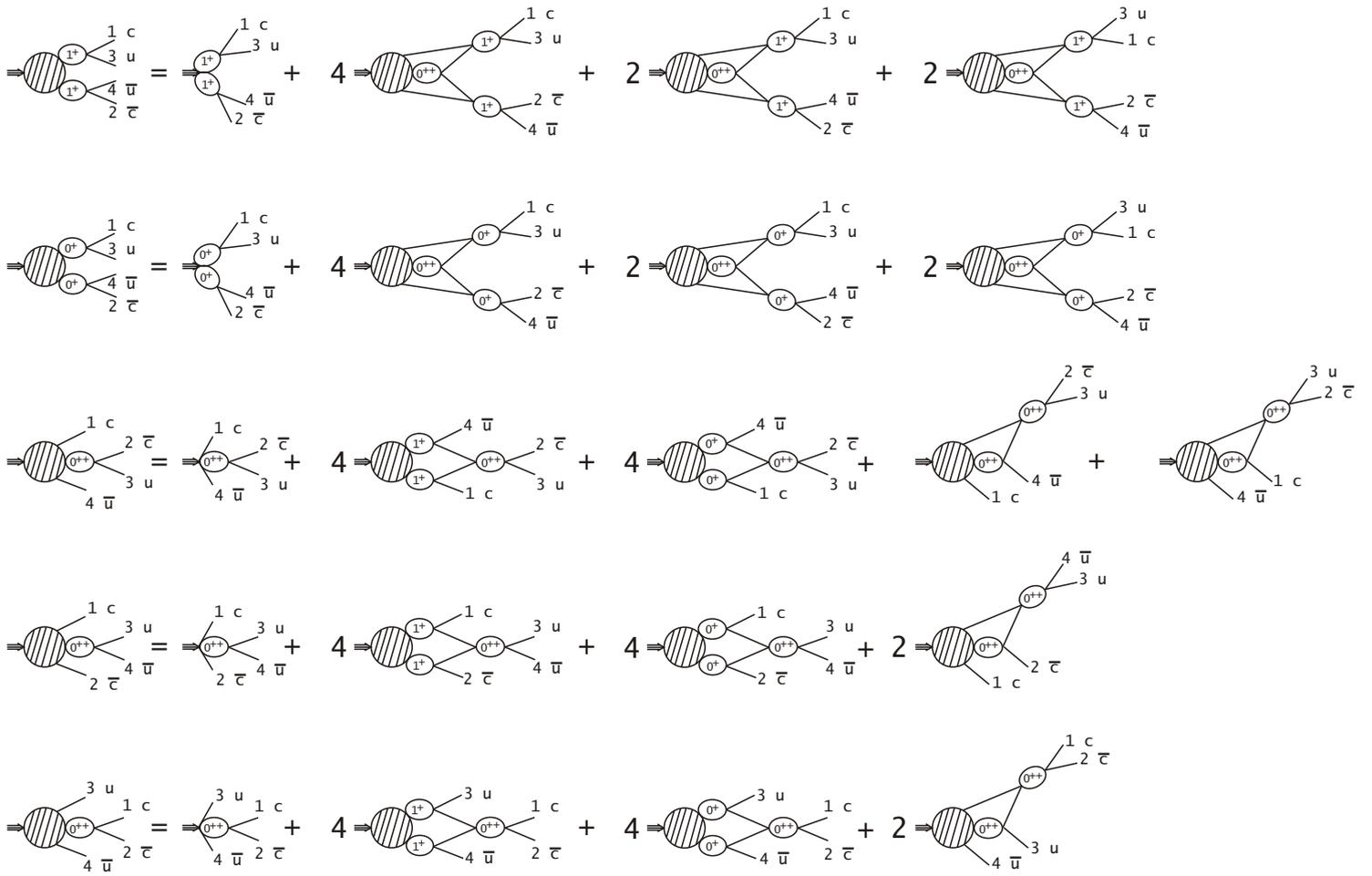

Fig. 4